\documentclass[]{spie}
\usepackage[]{graphicx}

\title{Laser Cooling of Dense Rubidium-Noble Gas Mixtures via Collisional Redistribution of Radiation}

\author{Ulrich Vogl\supit{a,$\dagger$}, Anne Sa{\ss}\supit{a} and Martin Weitz\supit{a}
\skiplinehalf
\supit{a} Institut f\"ur Angewandte Physik der Universit\"at Bonn, Wegelerstr. 8, 53115 Bonn, Germany
}

\authorinfo{ $\dagger$E-mail: uvogl@umd.edu\\
 Present address: Joint Quantum Institute, NIST and University of Maryland in College Park, USA}

\begin{document}
  \maketitle

\begin{abstract}
We describe experiments on the laser cooling of both helium-rubidium and argon-rubidium gas mixtures by collisional redistribution of radiation. Frequent alkali-noble gas collisions in the ultradense gas, with typically 200\,bar of noble buffer gas pressure, shift a highly red detuned optical beam into resonance with a rubidium D-line transition, while spontaneous decay occurs close to the unshifted atomic resonance frequency. The technique allows for the laser cooling of macroscopic ensembles of gas atoms. The use of helium as a buffer gas leads to smaller temperature changes within the gas volume due to the high thermal conductivity of this buffer gas, as compared to the heavier argon noble gas, while the heat transfer within the cell is improved.

\end{abstract}


\keywords{Laser cooling, optical refrigeration, collisional redistribution}

\section{INTRODUCTION}
\label{sec:intro}  
The idea to cool matter with light can be traced back at least 80 years \cite{Pringsheim}, but not until the invention of the laser viable proposals and experimental work has started. 
The greatest influence on further research in this area had the technique of Doppler cooling of atomic gases suggested by H\"ansch and Schawlow  \cite{Hansch}, which has lead to the extremely successful field of ultracold dilute atomic gases \cite{nobelchu,nobelcohen,nobelphillips}. In another approach for laser cooling, anti-Stokes processes in multilevel systems have been employed.
Proof of principle experiments for this type of laser cooling have been carried out with $CO_2$ gas \cite{djeu,bertolotti} and a dye-solution \cite{zander}. More recently, anti-Stokes laser cooling has allowed for the laser cooling of solid state materials, as rare-earth doped glasses \cite{mungan,clark}.
Cooling down to 150\,K starting from ambient temperature could be realized in such systems \cite{Sheik-Bahae2}.

In our work, we are examining laser cooling of an alkali-noble buffer gas mixture at very high pressures typically around a few hundred bar by means of collisional redistribution. As with the cooling of solids, this method involves the cooling of macroscopic ensembles of matter. Redistribution of atomic fluorescence is most widely known in the context of magnetooptic trapping of ultracold atoms, where this mechanism is a primary cause of trap loss processes \cite{adams}. In the long investigated field of room-temperature atomic collisions, redistribution of fluorescence is a natural consequence of line-broadening effects from collisional aided excitation \cite{yakovlenko}.
In theoretical works, Berman and Stenholm proposed in 1978 laser cooling (and heating) based on the energy loss or gain during collisionally aided excitation of atoms \cite{Berman_cool}. Experiments that used alkali vapor and moderate buffer gas pressure below one atmosphere observed a heating for blue detuned radiation \cite{Berman_giacobino}, but the cooling regime was never reached.

In our experiment laser cooling by collisional redistribution is investigated in a mixture of hot atomic rubidium vapor and several hundred bar of a buffer gas, where the collisionally broadened linewidth approaches the thermal energy $k_BT$ in frequency units. A scheme of the described laser cooling approach is sketched in Figure 1,
where the variation of the rubidium $
5S_{1/2}$ ground state and the $5P_{1/2}$ excited state versus the distance from a noble gas perturber atom is indicated in a binary collision quasimolecular picture.
During a collision between a rubidium and a noble gas atom, the rubidium electronic D-line transition can be transiently
 shifted into resonance with an otherwise far red detuned laser beam, which enables collisionally aided radiative excitation \cite{yakovlenko,  jablonski_prings,garbuny}. To achieve cooling, we typically work with detunings of 10-20\,nm to the red of an atomic resonance, where  excitation is energetically only possible with the support of a collisional process, and the excitation rate is strongly dependent on the rate of collisions.
\begin{figure}
	\centering
		\includegraphics[width=0.4\textwidth]{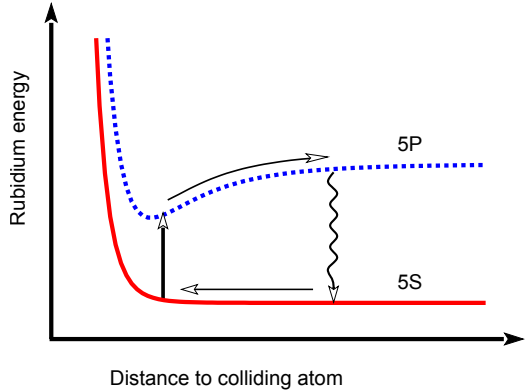}
	\caption{Cooling scheme. The atomic rubidium energy levels are shifted as a noble gas perturber atom approaches the optically active rubidium atom. Excitation with far red detuned light becomes possible during a collisional process.}
	\label{fig:setnatger}
\end{figure}
The duration of a collision and the excited state lifetime are clearly resolved ($\approx$ps compared to 27\,ns), so after an excitation, spontaneous decay occurs most likely at larger distances between the collision partners and thus closer to the unpertubed atomic resonance frequency. On average the emitted photon is blue shifted relatively to the incident wavelength. The rubidium atoms undergo frequent cycles of collisionally aided excitation and spontaneous decay, and provided that the emitted photons can leave the sample, the collision partners loose kinetic energy and will be cooled. The typical energy that can be removed from the reservoir in a cooling cycle is of order $k_BT\simeq\hbar\cdot 10^{13}$\,Hz at T=500\,K, corresponding to a factor 20000 above the typical energy difference $\hbar k\cdot v$ removed by a fluorescence photon in usual Doppler cooling of dilute rubidium atomic gases.

Experimentally, we have in earlier work demonstrated relative cooling of a rubidium-argon mixture by 66\,K and of a potassium-argon mixture by 132\,K, both with an argon buffer gas pressure near 200\,bar \cite{cooling,sass}. The achieved cooling power approaches 100\,mW, and the achieved temperature drop is limited by the thermal conductivity of the gas. The partial pressure of the buffer gas exceeds that of the alkali vapor by five orders of magnitude, and correspondingly it 
are the thermal conductivity and the thermal diffusivity of the buffer gas that determine the induced temperature change and the temperature flow from a local heat sink in the gas. Accordingly it is favourable to use a buffer gas with low thermal conductivity to achieve a large temperature gradient and a low local minimum temperature in the gas. On the other hand, for the purpose of actually transferring the cooling power to another cooling load which should not be directly exposed to the laser radiation, the exact opposite is a good choice. Here we present experimental results where we used as buffer gas helium, the noble gas with the highest thermal conductivity and thermal diffusivity. Helium has in both parameters roughly an order of magnitude higher values than argon, the  buffer gas used in our previously published results \cite{cooling,sass}.

The present paper describes spectroscopic measurements and initial redistribution cooling results using helium as a buffer gas. We are aware that the achieving of larger temperature drops in helium-based buffer gas systems will most likely require the use of thermally isolated all-sapphire cells.
Moreover, we review earlier experiments obtained with rubidium-argon buffer gas mixtures.

In the following, section\,2 describes the used experimental setup and measurements of the buffer gas broadened sample. Further, section\,3 describes results on redistribution laser cooling and section\,4 gives conclusions.

\section{Experimental setup and characterization of the pressure broadened system}
A scheme of the used experimental setup is shown in Fig.\,2, see \cite{JMO} for a detailed description. In brief, the present experiment is carried out in a high pressure steel chamber, where optical access is provided by two uncoated sapphire optical windows. After ultra-sound cleaning and bake-out, the cell is filled with rubidium metal (1\,g ampules) and helium or argon buffer gas. The cell is heated to typically 500\,K, which yields a vapor pressure limited rubidium partial pressure of $\approx$1\,mbar, corresponding to a number density of  $\approx$10$^{16}$cm$^{-3}$, and the used noble gas pressure is typically around 200\,bar.
 As primary laser source for spectroscopy and cooling we use a continuous-wave titanium-sapphire laser system of 3\,W output power near the wavelength of the rubidium D-lines.
 
 For spectroscopic characterization of the pressure broadened sample, a confocal detection geometry is used (see also \cite{pra}), to suppress scattered incident laser radiation and selectively detect fluorescence from the cooling beam region. Figure\,3a gives a typical pressure broadened spectrum. The plot shows the total detected fluorescence yield versus the incident laser frequency in a region around the rubidium D1- and D2-lines (at 377\,THz and 384\,THz, corresponding to 795\,nm and 780\,nm respectively). 
\begin{figure}
	\centering
		\includegraphics[width=0.5\textwidth]{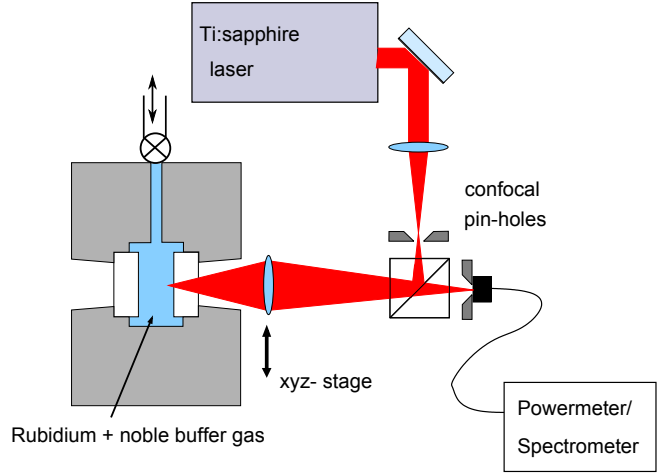}
	\caption{Experimental setup used for fluorescence spectroscopy of the dense buffer gas broadened rubidium ensemble. To suppress a background of scattered light, a confocal detection method is used.}
	\label{fig:setnatger}
\end{figure}
 The lineshape is asymmetric, with an increased absorption amplitude on the blue wing of the line for the here shown case of the use of helium as a buffer gas. The large pressure broadening enables to efficiently excite the rubidium atoms at the large detuning values of order $k_BT/\hbar$ (corresponding to 23\,nm at T=500\,K) as typically used in the cooling experiment.
\begin{figure}
	\centering
		\includegraphics[width=0.8\textwidth]{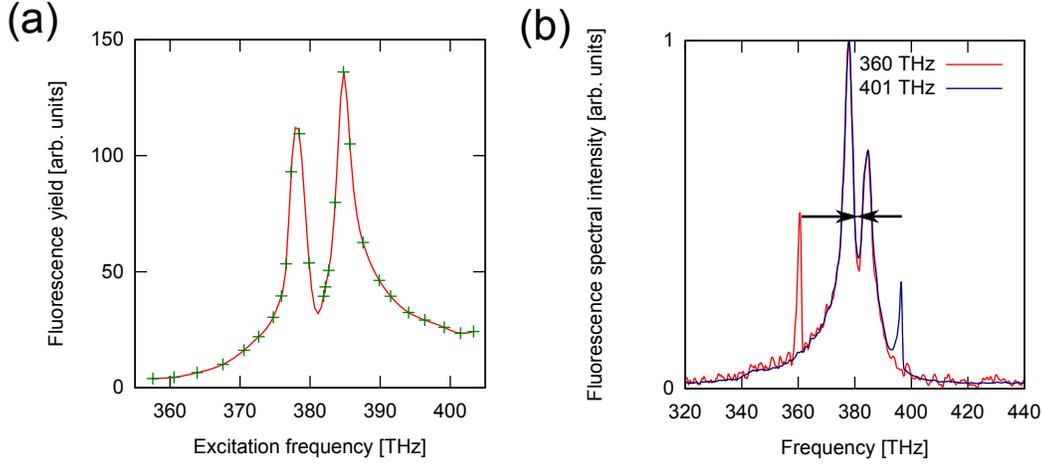}
	\caption{(a) Measured fluorescence power for atomic rubidium vapor at 530\,K temperature with 200\,bar pressure of helium buffer gas, versus the excitation wavelength.
(b) Spectrally resolved fluorescence signal of rubidium vapor at 530\,K with 200\,bar helium buffer gas. Shown are cases for one red detuned (360\,THz) and one blue detuned optical excitation frequency (401\,THz). The shown arrows indicate the experienced frequency shift of the scattered photons by the redistribution process. Note that both spectra are normalized and that at excitation frequency 360\,THz the total fluorescence power is significantly lower than for 401\,THz, following the behaviour shown in panel (a).}
	\label{fig:specboth}
\end{figure}

Figure 3b shows spectrally resolved fluorescence spectra for exciting laser frequencies of 360\,THz and 401\,THz, corresponding to a red and a blue detuning respectively.
In both cases, the scattered fluorescence is efficiently redistributed towards the D-lines center. While for the former case the emitted fluorescence has (in average) higher photon energies than the incident radiation, which we expect to lead to cooling of the gas sample, in the latter case the emitted fluorescence has lower photon energies than the exciting optical beam, for which heating is expected.

Provided that no additional heating processes occur, we from energy conservation expect a cooling power in the sample of
\begin{equation}	 P_{\mathrm{cool}}=P_{\mathrm{opt}}a({\nu})\frac{\overline{\nu_{\mathrm{fl}}}-\nu_{\mathrm{laser}}}{\nu_{\mathrm{laser}}},
\label{eq:pcool}
\end{equation}
where $P_{\mathrm{opt}}$ denotes the incident laser power and $a(\nu)$ the absorption probability in the gas for an incident laser frequency $\nu$. Further,  $\nu_{\mathrm{fl}}$ is the centroid frequency of the emitted fluorescence.
The frequency dependence of the optical density follows in lowest order the fluorescence spectrum
shown in Figure 3a and on resonance reaches a measured maximum value of about 4.5.
The expected cooling power, according to Eq.\,1, versus the laser frequency is shown by the connected red crosses in Figure 5a.
 For a red detuning of the laser light of 10-20\,nm from resonance 
 the expected cooling power for the here investigated helium-rubidium gas mixture is
 in the order of 5 mW. This is significantly lower than in results obtained for the case of an argon buffer gas in earlier measurements of our group  \cite{cooling,sass}, for which the obtained cooling power was roughly an order of magnitude higher (see connected red crosses in Fig.\,5b) at comparable laser power. Besides of the lower rubidium vapor pressure for the shown data set, we also attribute this to the different quasi-molecular potentials of the Rb-He and Rb-Ar pairs, which result in a lower absorption for far red detuned radiation for the Rb-He case.
 \newline

\section{LASER COOLING OF THE GAS MIXTURE}

To obtain evidence for an actual temperature change of the gas inside the cell, we use a thermal deflection spectroscopy technique \cite{Boccara,spear,whinnery}, see Fig.\,4. The cooling laser beam induces a local temperature change in the gas, which in turn changes the refractive index of the gas. We detect this change of the refractive index profile with an off-resonant probe laser beam, in this case a helium-neon laser at 632\,nm. The 
induced change of the refractive index $n$ versus temperature can be approximated as
\begin{equation}
	\frac{dn}{dT}\simeq-\frac{n-1}{T}.
\end{equation}
\begin{figure}
	\centering
		\includegraphics[width=0.6\textwidth]{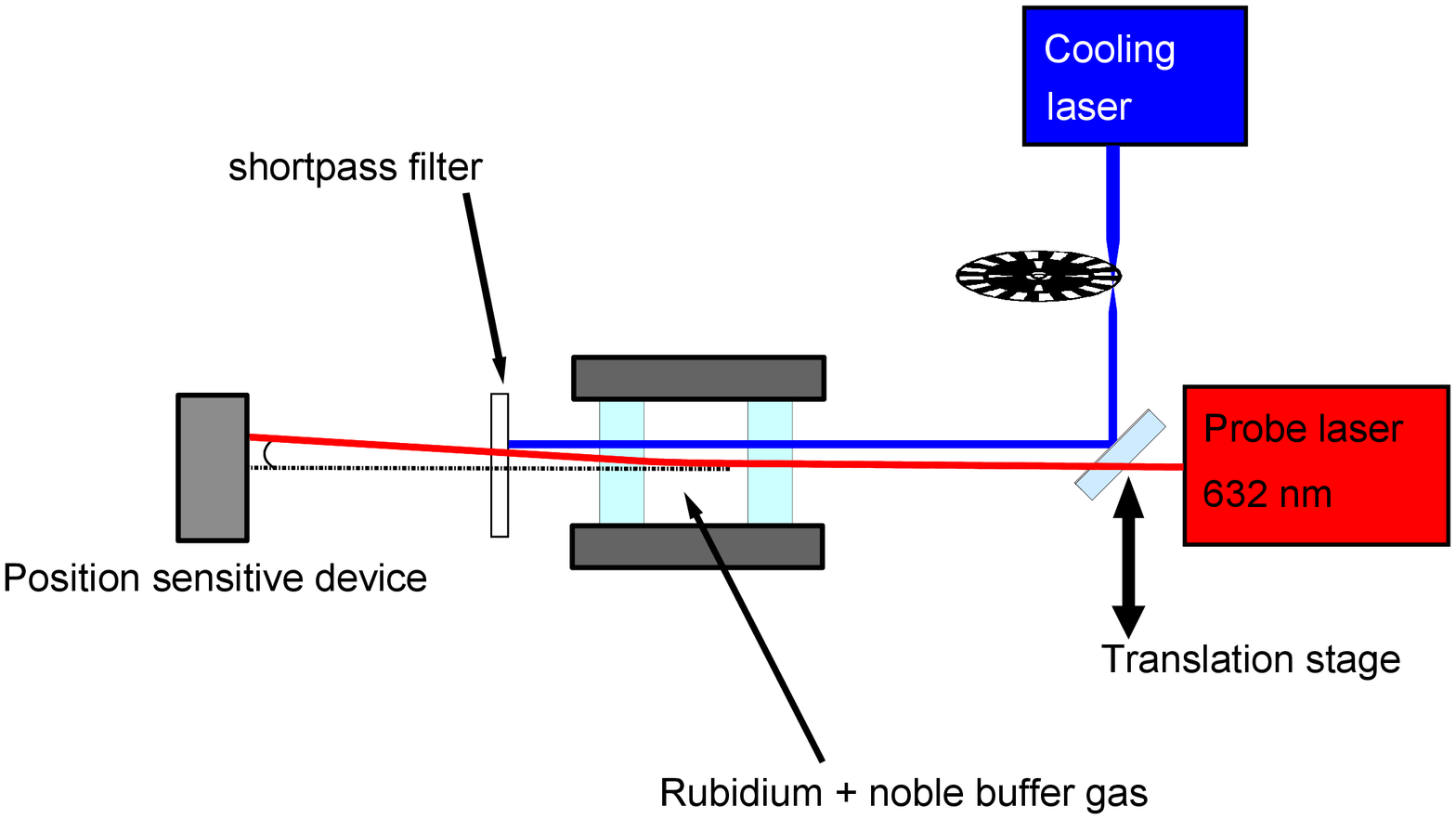}
	\caption{Experimental setup used in the thermal deflection spectroscopy measurements. To determine the deflection angle of the probe laser beam, either a split photodiode or a position sensitive device was used. }
	\label{fig:setnatger}
\end{figure}
The deflection of the probe beam along the beam path is accordingly given by
\begin{equation}
	\Phi=\frac{n-1}{T}\int^{L}_{0}\frac{dT(r,z)}{dr}dz.
\end{equation}
The cooling beam has a $TEM_{00}$ profile with beam radius $w$ and experiences an absorption coefficient $\alpha$.
For a heat transport mainly in the radial direction (we assume a long cell and that the beam focal radius is much smaller than the absorption length), and assuming that the cooling beam removes heat with a rate following its intensity distribution, the temperature distribution can be determined from the beam deflection using:
\begin{equation}
	\Delta T(r,z)=\frac{T}{n-1}\frac{\alpha e^{(-\alpha z)}}{1-e^{(-\alpha z)}}\int^{\infty}_{r}\Phi(r')dr',
\end{equation}
where $r$ denotes the transverse offset between cooling and probe laser beams. For the refractive index of the Rb-He mixture at the 632\,nm probe wavelength a value n$\approx$1.0070(3) for helium at 200\,bar is used \cite{born}, where the quoted uncertainty is basically determined by the accuracy of the pressure measurement.

The temperature gradient for the given situation can be modelled using from a heat transfer model \cite{whinnery}. After a cooling time $t$, the expected temperature gradient profile is
\begin{equation}
	\frac{dT}{dr}=-\frac{P_{\mathrm{cool}}}{2\pi\kappa}\frac{e^{-\alpha z}}{r}\left[\exp\left(\frac{2r^2}{w^2}\right)-\exp\left(\frac{2r^2}{w^2-8Dt}\right)\right].
\label{eq:four2}
\end{equation}
where $\kappa$ is the specific heat capacity, $w$ the beam radius (1\,mm for both cooling and probe beam),  $D$ the thermal diffusivity, and $P_{\mathrm{cool}}$ is the cooling power.

\begin{figure}
	\centering
		\includegraphics[width=0.95\textwidth]{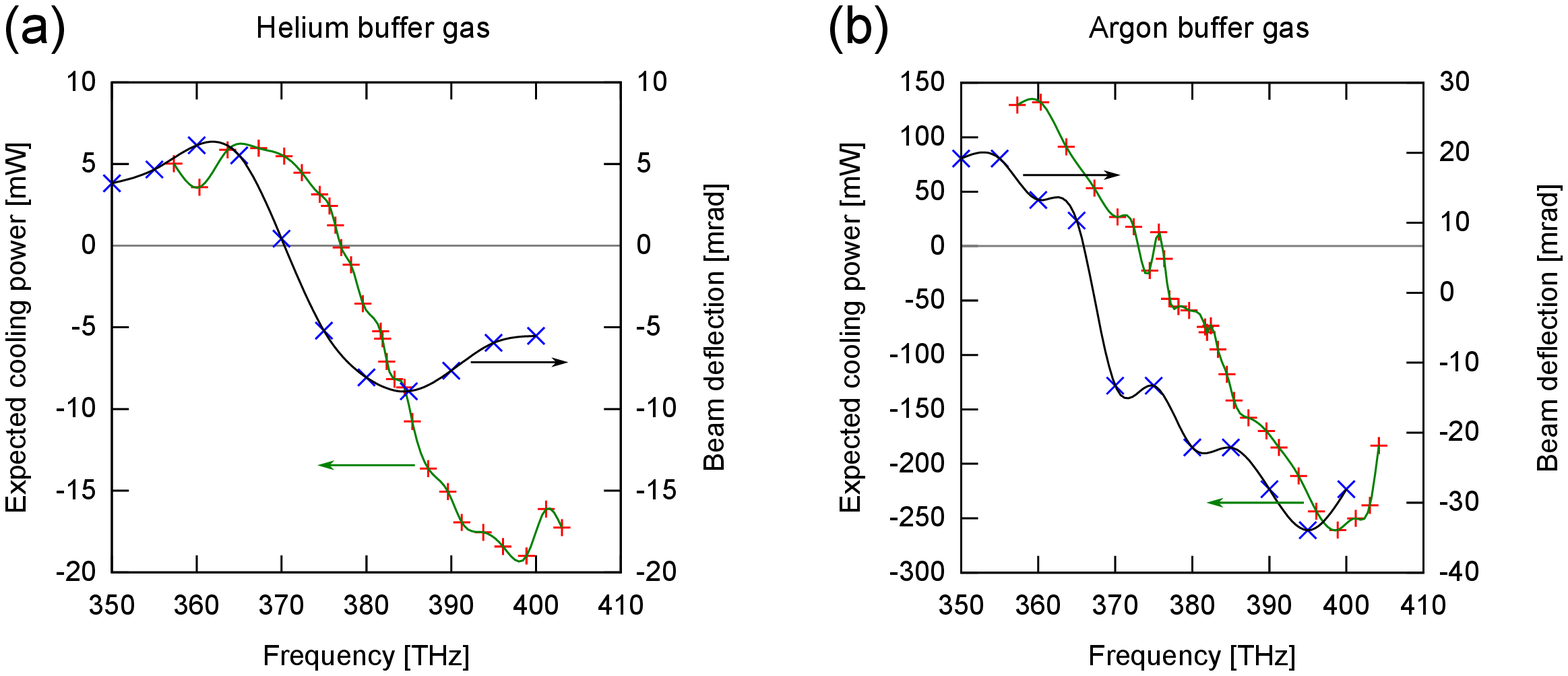}
	\caption{A comparison between the expected cooling power (connected red crosses) and the observed deflection (connected blue crosses) on the induced thermal lens for (a) the buffer gases helium (200\,bar) and (b) the values obtained earlier for argon (230\,bar) \cite{cooling}.  }
	\label{fig:setnatger}
\end{figure}

The connected blue crosses in Figs.\,5a and 5b show the measured deflection angle for helium and argon buffer gas respectively
 as a function of the cooling-laser frequency. For red detuning the probe beam is deflected towards the cooling beam, which indicates cooling, while the probe beam is deflected away from the cooling beam when the gas is locally heated. With the thermal deflection technique for both measurements cooling is observed only  when the laser frequency is detuned approximately 10\,THz lower than one would expect from the redistribution of the fluorescence. The reason for this can be residual heating, which could be due to a small quenching cross section of excited-state rubidium atoms under the here present high buffer gas pressure. Direct measurement of the lifetime of the excited 5P rubidium levels under 200 bar helium buffer gas confirm the natural lifetime of 27\,ns within an experimental uncertainty of 5\,ns \cite{igor}. The here observed shift could be explained by an average decrease of the excited state lifetime of the rubidium 5P state due to quenching  by roughly 1\,ns. 
Despite the use of high purity buffer gas (we typically use helium and argon buffer gas of purity 6.0), we cannot exclude that residual impurities, for example by a small nitrogen admixture, result in a corresponding reduction of the lifetime due to the high number of collisions the rubidium atoms undergo, typically several 10000 within a natural lifetime.

\begin{figure}
	\centering
		\includegraphics[width=0.6\textwidth]{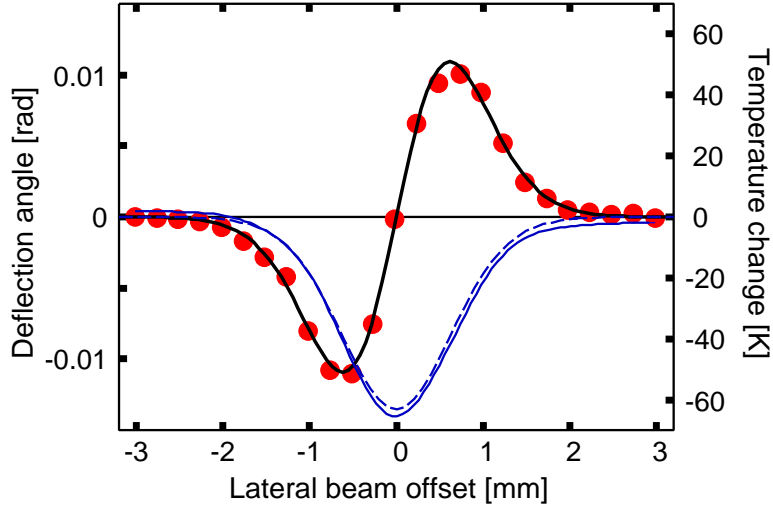}
	\caption{The red dots show results for probe-beam angular deflection as a function of lateral offset between the probe and the cooling beam, and the black line gives a fit to a theoretical model. With the deflection measurement we sample the radial temperature gradient induced by the cooling beam. The solid blue line shows the corresponding temperature profile near the cell entrance. The determined temperature drop in the beam center is 66(13)\,K. The dashed blue line follows a heat-transport model (Eq.\,4). Reproduced from Ref.\,18.  }
	\label{fig:setnatger}
\end{figure}
Quantitative measurements of the obtained temperature profile induced by the cooling laser beam have been carried out for the case of the rubidium-argon mixture, see Ref.\,\cite{cooling}. Here the lateral offset between the cooling and the probe laser beam was scanned. The red dots fitted with the black line show corresponding deflection data, obtained in such a transversal scan, which is a measurement for the transverse temperature gradient. The blue line shown in the figure is the result of a numerical integration of the corresponding deflection data following Eq.\,4. The used parameters for this measurement were a cooling laser frequency of 365\,THz, 1\,mm cooling and probe beam diameters respectively, and 90\% optical absorption in the 1\,cm long cell. The obtained temperature drop in the beam center is 66(13)\,K, which is attributed to be limited by thermal conductivity of the argon gas in the thermally not isolated cooling volume. The dashed blue line is the temperature profile obtained from a theoretical heat transport model from which a cooling power of 87(20)\,mW can be derived. We expect that the temperature drop can be increased by depositing the available cooling power into a smaller volume, as can be achieved with a gas of higher optical density.

\section{Conclusion}

We have investigated collisional redistribution cooling in both rubidium-argon and rubidium-helium gas mixtures. The observed cooling performance is, due to both the smaller thermal conductivity of the argon buffer gas and a blue asymmetry of the spectra in the case of rubidium-helium pair, generally more favourable in case of the use of argon as a buffer gas, at least for the present experimental condition of a gas cooling within the laser focus alone. Cooling limitations from the large thermal conductivity of the buffer gas can however be avoided, when cooling of the complete cell is achieved, which requires the use of all (sapphire-) glass cells to suppress heating from wall absorption of scattered fluorescent radiation. To avoid condensation of the alkali atoms at the cell walls at low temperatures, we have started to investigate mixtures of molecular gases, which can be gaseous at room temperatures, and noble gases. Promising candidates for a molecular redistribution cooling (besides the alkaline-earth monohydrides, for which Doppler laser cooling has already been demonstrated \cite{2010demille}, are e.g. ethylene or CO$_2$-gas. Future applications of redistribution laser cooling can include the supercooling of gases below the homogeneous nucleation temperature \cite{Kivelson, cavagna,Simeoni} and optical chillers \cite{Sheik-Bahae2}.

 \acknowledgments     
Financial support
from the Deutsche Forschungsgemeinschaft is acknowledged.


\end{document}